\pdfoutput=1
\documentclass[10pt,letterpaper,twoside,notitlepage]{article}
\usepackage{amsfonts}
\usepackage[intlimits]{amsmath}
\usepackage{amssymb}
\usepackage{amsxtra}
\usepackage[dvipsnames]{xcolor}
\usepackage{colordvi}
\usepackage{colortbl}
\usepackage{float}
\usepackage{floatflt}
\usepackage{graphicx}
\usepackage{hhline}
\usepackage[compat=1.1.0]{tikz-feynhand}
\usepackage{xspace}
\usepackage[bookmarks=true,%
				bookmarksnumbered=true,%
				bookmarksopen=true,%
				allbordercolors=lime!60!white]%
				{hyperref}
\providecommand{\nc}{\newcommand}
\nc{\redcol}{\color{red!80!black}\,}
\nc{\vercol}{\color{green!24!black}\,}
\nc{\blucol}{\color{blue!40!black}\,}
\nc{\brocol}{\color{red!30!black}\,}
\nc{\txcol}{\color{black}\,\xspace}

\nc{\itembf}[1][+]{\item[\textbf{#1}]}

\numberwithin{figure}{section}  		

\usepackage{geometry}
\begin{document}%
%
%
\thispagestyle{empty}
\begin{centering}
	\Large\bfseries%
	TikZ-FeynHand: Basic User Guide%
	\vspace{1mm}\\
	\large\mdseries%
	Max Dohse%
	\vspace{1mm}\\
	\normalsize%
	\today
	\vspace{1mm}\\
	\small\itshape
	Written at Instituto de F\'isica y Matem\'aticas (IFM-UMSNH),
	Universidad Michoacana de San Nicol\'as de Hidalgo,\\
	Edificio C-3, Ciudad Universitaria, 
	58040 Morelia, Michoac\'an, M\'exico.
	\vspace{1.5mm}\\
\end{centering}
%
%
\hrule
\vspace{3mm}
{\centering\Large\bfseries Quick Reference\\}
\noindent
\textbf{Environment}\\
\vercol\verb!\begin{feynhand}... \end{feynhand}!\txcol
enclosed in \vercol\verb!\begin{tikzpicture}[OPTIONS] ... \end{tikzpicture}!\txcol\\
OPTIONS: \vercol\verb!baseline=1cm, baseline=(v.base), node distance = 2cm!\txcol,
no options are mandatory.
\vspace{1.5mm}\\
\noindent
\textbf{Vertices}\\
\blucol\verb!\vertex (v0) at (x0,y0);!\txcol\\
\blucol\verb!\vertex [particle] (v1) at (x1,y1) {e$^-$};!\txcol\\
\blucol\verb!\vertex [STYLE] (v2) at (x2,y2)! \redcol\verb!{}!\blucol\verb!;!\txcol\\
STYLE options: \blucol\verb!dot, ringdot, squaredot, crossdot, blob, blobring, blobgray, blobNW, blobNE!\txcol\\
If STYLE option omitted/empty: nothing is drawn, vertex only defined for later use (drawing propagators).
\vspace{1.5mm}\\
\noindent
\textbf{Propagators}\\
\vercol\verb!\propag [STYLE] (v1) to [EDGE] (v2);!\txcol\qquad\qquad
(If STYLE option omitted/empty: straight line is drawn.)\\
\vercol\verb!\graph {(v1) --[STYLE, EDGE] (v2) --[STYLE, EDGE] (v3) --[STYLE, EDGE] (v4)};!\txcol
\\[1mm]
Short STYLE options: \vercol\verb!fer, antfer, pho, bos, chabos, antbos, sca, chasca, antsca,!\\ 
\textcolor{white}{Short STYLE options:} 
\verb!glu, gho, chagho, antgho, maj, antmaj, plain, mom, revmom!\txcol\\
EDGE options: \vercol\verb!edge label, in, out, looseness, half left, quarter right!\txcol
\\
Crossing propagators:\quad
STYLE option \redcol\verb!top!\txcol, 
for example: \vercol\verb!\propag[fer,! \redcol\verb!top!\vercol\verb!] (a) to (b);!\txcol
\vspace{1.5mm}\\
\noindent
\textbf{Customize}\\
\blucol\verb!\setlength{\feynhanddotsize}{LENGTH}!\txcol\\
analogue: 
\blucol\verb!\feynhandblobsize, \feynhandlinesize, \feynhandarrowsize, \feynhandtopsep!\txcol\\
\blucol\verb!\renewcommand{\feynhandtopsepcolor}{COLOR}!\txcol\\
\blucol\verb!\tikzfeynhandset{every STYLE={/tikz/color=COLOR},}!\txcol\\
\vspace{1mm}
%
\hrule
\vspace{2mm}

{\centering Version 1.0.0 - 
 available at\qquad  \texttt{https://ctan.org/pkg/tikz-feynhand}\\}

\vspace{1mm}

\hrule
%
\tableofcontents
%
%
%
%
\section{Goals and related software}
\label{sec:_goals}
\noindent
\textbf{TikZ-FeynHand} is a low-end version of Joshua Ellis' original
\textbf{TikZ-Feynman} package \cite{tikzfeynman} (Version 1.1.0),
which has provided efficient new ways of drawing beautiful Feynman diagrams automatically 
(and manually).
FeynHand is a modification for those users who want \textbf{easy access} to TikZ-Feynman's
drawing styles \textbf{in LaTex without any reference to LuaTex},
and who want to \textbf{customize the diagrams} more easily.
FeynHand thus aims to be an alternative way to access
the manual mode of its big brother TikZ-Feynman.

The drawing styles of TikZ-FeynHand are available through simple commands 
following the \textbf{syntax conventions of TikZ,} 
which uses Tex-like commands to create graphics.
Since there are still \textbf{journals} which do not yet support TikZ in submissions,
TikZ is capable of creating separate PDF graphic files,
which can then be submitted with the article.
This is called \textbf{externalizing the graphics} and is outlined in
Appendix~\ref{Apx_tikzbasics_externalize},
see also Section~50.4 in the TikZ/PGF Manual~\cite{tikz}.
\\[1.5mm]
\noindent
The \textbf{GENERAL WORKFLOW} for creating a diagram with TikZ-FeynHand is the following: 
\begin{itemize}
\itembf[0.]
	Initialize the \vercol\verb$feynhand$\txcol environment 
	inside the \vercol\verb$tikzpicture$\txcol environment\\
	(in turn typically within a \vercol\verb$figure$\txcol or some math environment).
	\itembf[1.]
	Set the positions of the vertices and the styles they are drawn in.
	\itembf[2.]
	Insert the propagator lines between the vertices.
	\itembf[x.]
	Draw extra objects (like coordinate systems).
\end{itemize}
FeynHand offers additonal shorter versions of TikZ-Feynmans TikZ-keys,
because I find some of them quite long, e.g.~\vercol\verb$anti charged boson$\txcol.
With little practice and efficient use of copy and paste, 
the ease of creating diagrams by hand is similar in both packages.
FeynHand contains a few tweaks that are handy for me and might be for others. 
\\[1mm]
The other main \textbf{features of FeynHand that differ from TikZ-Feynman} 
(Version 1.1.0) are listed below.
\begin{itemize}
	\item
	Package does neither load the Tex package \vercol\verb$ifluatex$\txcol,\\
	nor TikZ-Feynman's Lua patches \vercol\verb$tikzfeynman.patch.x.x.x.lua$\txcol.
	\item
	No extra \vercol\verb$\diagram$\txcol environment is needed, 
	since everything is placed by hand.
	\item
	\vercol\verb$top$\txcol style option for crossing propagators introduced, 
	see Section~\ref{sec:_feynmandiagrams_propagators_top}.
	\item
	Some additional vertex and propagator styles are available,
	see Sections~\ref{sec:_feynmandiagrams_vertices} 
	and \ref{sec:_feynmandiagrams_propagators}. 
	\item
	Shortcut commands for quickly customizing the diagrams' look,
	see Section~\ref{sec:_feynmandiagrams_customizing}.
	\item
	For brevity, we access/leave math mode via \$ instead of the \vercol\verb$\($\txcol 
	and \vercol\verb$\)$\txcol suggested in TikZ-Feynman's manual.\\
	However, this is rather a personal preference and the opinions on this topic are divided.
\end{itemize}
This user guide just describes briefly \textbf{how to quickly use FeynHand.}
There is also a three-page introduction to TikZ in Appendix~\ref{Apx_tikzbasics}.
For more information on TikZ-Feynman's capabilities and commands, 
we recommend its elaborate documentation \cite{tikzfeynman}, 
in particular Section~3, and also recommend the introduction 
and extensive manual of TikZ \cite{tikz}.
In all commands and keys mentioned in \cite{tikzfeynmanArticle}
and the corresponding files, the string \vercol\verb$feynman$\txcol has been replaced
by \vercol\verb$feynhand$\txcol in the FeynHand package,
in order to prevent any clash of variable names, 
if both TikZ-Feynman and FeynHand are used together in the same LaTex document.
%
%
\subsection*{Acknowledgments}
\noindent
I am grateful to Joshua Ellis for friendly advice 
and for writing TikZ-Feynman in the first place.\\
I am also thankful to Till Tantau and all collaborators for writing TikZ/PGF.
%
%
%
\subsection*{Citations}
\noindent
If you use TikZ-FeynHand for any publication, please be so kind to\\
cite both Joshua Ellis' original manual \cite{tikzfeynmanArticle} as well as this userguide.
%
%
\subsection*{License}
\noindent
This PDF and the whole package are free: you can redistribute it and/or modify it under the
terms of the GNU General Public License as published by the Free Software
Foundation, either version 3 of the License, or (at your option) any later version.
%
%
\newpage
\section{Feynman diagrams with TikZ-FeynHand}
\label{sec:_feynmandiagrams}
%
%
\subsection{Environment}
\label{sec:_feynmandiagrams_environment}
\noindent
The \blucol\verb$feynhand$\txcol package is loaded as usual
with \blucol\verb$\usepackage{tikzfeynhand}$\txcol.
All FeynHand commands are used within the \blucol\verb$feynhand$\txcol
environment, which needs to be placed inside the \blucol\verb$tikzpicture$\txcol
environment%
\footnote{It is possible to define \texttt{feynhand} such that it includes \texttt{tikzpicture}. 
	We decide against this so that \texttt{feynhand} 
	can be used as a scope inside \texttt{tikzpicture}.
	Further, doing so would make externalizing the graphics more difficult.}.
Of course, all \textbf{TikZ commands remain available} within \blucol\verb$feynhand$\txcol,
and TikZ commands can also be placed outside of \blucol\verb$feynhand$\txcol
inside the \blucol\verb$tikzpicture$\txcol enclosing it.

The \textbf{coordinate system} used by TikZ is the usual cartesian (x,y)-plane,
with the x-axis pointing horizontally to the right
and the y-axis pointing vertically upwards on the page.
The default coordinate unit is centimeters.
We recall that all TikZ and/or FeynHand commands must \textbf{end with
a \;;\; (semicolon).}

When we want to use captions and labels, we enclose \blucol\verb$feynhand$\txcol 
in the \blucol\verb$figure$\txcol environment as below.
\blucol\begin{verbatim}
\begin{figure}
   \centering
   \begin{tikzpicture}
      \begin{feynhand}

         ... FeynHand and/or TikZ commands ...

      \end{feynhand}
   \end{tikzpicture}
   \caption{A diagram drawn with TikZ.}
   \label{Fig:_diagram1}
\end{figure}
\end{verbatim}\txcol
%
\vspace{-5mm}
\subsection{Equations involving Feynman diagrams}
The \vercol\verb$feynhand$\txcol environment can also be used directly
in math environments. For example, the code below produces the (nonsense) equation below it.
For the \vercol\verb$baseline$\txcol option of the
\vercol\verb$tikzpicture$\txcol environment, see Section~12.2.1
in the TikZ/PGF Manual \cite{tikz}.
In the first diagram we have set it to the y-coordinate of the vertex (o),
and in the second we have set it to the value y=-0.3cm.
That is, in each diagram the corresponding horizontal lines of constant y-coordinate 
are aligned with the baseline of the math environment.
\vercol\begin{verbatim}
\begin{align*}
   \int dx\; f(x) = \alpha
   \begin{tikzpicture}[baseline=(o.base)]
      \begin{feynhand}
         \vertex (a) at (-1,-1); \vertex (b) at (1,-1); \vertex (c) at (0,1); 
         \vertex [dot, Blue] (o) at (0,0) {}; \propag [fermion, Blue] (a) to (o);
         \propag [anti fermion, Blue] (b) to (o); \propag [fermion, Blue] (c) to (o);
      \end{feynhand}
   \end{tikzpicture}
   - 2i\,e
   \begin{tikzpicture}[baseline=-0.3cm]
      \begin{feynhand}
         \vertex (a) at (-1,-1); \vertex (b) at (1,-1); \vertex (c) at (0,1); 
         \vertex [dot, Orange] (o) at (0,0) {}; \propag [photon, Orange] (a) to (o);
         \propag [photon, Orange] (b) to (o); \propag [photon, Orange] (c) to (o);
      \end{feynhand}
   \end{tikzpicture}
\end{align*}
\end{verbatim}\txcol
\vspace{-13mm}
\begin{align*}
   \int\!\! dx\; f(x) = \alpha
   \tikzsetnextfilename{baseline_obase}
   \begin{tikzpicture}[baseline=(o.base)]
      \begin{feynhand}
         \vertex (a) at (-1,-1); \vertex (b) at (1,-1); \vertex (c) at (0,1); 
         \vertex [dot, blue] (o) at (0,0) {}; \propag [fermion, Blue] (a) to (o);
         \propag [anti fermion, Blue] (b) to (o); \propag [fermion, Blue] (c) to (o);
      \end{feynhand}
   \end{tikzpicture}
   - 2i\,e
   \tikzsetnextfilename{baseline_03cm}
   \begin{tikzpicture}[baseline=-0.3cm]
      \begin{feynhand}
         \vertex (a) at (-1,-1); \vertex (b) at (1,-1); \vertex (c) at (0,1); 
         \vertex [dot, orange] (o) at (0,0) {}; \propag [photon, Orange] (a) to (o);
         \propag [photon, Orange] (b) to (o); \propag [photon, Orange] (c) to (o);
      \end{feynhand}
   \end{tikzpicture}
\end{align*}
%
%
\subsection{Vertices}
\label{sec:_feynmandiagrams_vertices}
\noindent
The \vercol\verb$\vertex$\txcol command is the same as in TikZ-Feynman,
and essentially consists of the command \vercol\verb$node$\txcol of TikZ.
By declaring a vertex, we mean defining its name and its position 
in the coordinate system, and specifying how the vertex is to be drawn, 
see below for examples.
The vertices' names are mere labels for later reference
(just like equation labels in Latex)
and don't show up anywhere in the graph.
The idea is now to use the vertices' names as a reference for their  
coordinates. This makes their use and modification much easier.
For example, we can draw a line from vertex (a) to (b) by
\vercol\verb$\draw (a) to (b);$\txcol 
see Section~\ref{sec:_feynmandiagrams_propagators}.

Below we list the available vertex styles.
Some users like to draw a gap in propagator lines which cross each other
without any interaction being implied there.
This can be done using gap vertices, 
see Section~\ref{sec:_feynmandiagrams_propagators_top}.
%
\subsubsection*{Bare vertex}
\noindent
This is the basic vertex: nothing is drawn at the vertex itself,
it merely connects propagator lines.
Note that we do not put an empty argument \redcol\verb${}$\txcol at the end.
\vspace{2mm}\\
\begin{minipage}{0.7\linewidth}
\vercol\begin{verbatim}
\begin{tikzpicture}
   \begin{feynhand}
   \vertex (a) at (0,0);      \vertex (b) at (2,0);
   \vertex (c1) at (4,0.5);   \vertex (c2) at (4,-0.5);
   \propag [plain] (a) to (b);
   \propag [plain] (b) to (c1);
   \propag [plain] (b) to (c2);
   \end{feynhand}
\end{tikzpicture}
\end{verbatim}\txcol
\end{minipage}
\begin{minipage}{0.25\linewidth}
\tikzsetnextfilename{vertex_bare}
\begin{tikzpicture}
   \begin{feynhand}
   \vertex (a) at (0,0);
   \vertex (b) at (2,0);
   \vertex (c1) at (4,0.5);
   \vertex (c2) at (4,-0.5);
   \propag [plain] (a) to (b);
   \propag [plain] (b) to (c1);
   \propag [plain] (b) to (c2);
   \end{feynhand}
\end{tikzpicture}
\end{minipage}
%
\subsubsection*{Particle vertex}
\noindent
Particle vertices are usually used only for incoming 
and outgoing particles. \\
For intermediate particles, see propagator edge labels in 
Section~\ref{sec:_feynmandiagrams_proplabelsmomentum}.
\vspace{2mm}\\
\begin{minipage}{0.7\linewidth}
\vercol\begin{verbatim}
   ...
   \vertex [particle] (a) at (0,0) {e$^-$};
   ...
\end{verbatim}\txcol
\end{minipage}
\begin{minipage}{0.25\linewidth}
\tikzsetnextfilename{vertex_particle}
\begin{tikzpicture}
   \begin{feynhand}
   \vertex [particle] (a) at (0,0) {e$^-$};
   \vertex (b) at (2,0);
   \vertex (c1) at (4,0.5);
   \vertex (c2) at (4,-0.5);
   \propag [plain] (a) to (b);
   \propag [plain] (b) to (c1);
   \propag [plain] (b) to (c2);
   \end{feynhand}
\end{tikzpicture}
\end{minipage}
%
\subsubsection*{Dot vertex}
\noindent
Some users like to place dots on vertices in order to emphasize
that an interaction occurs there.\\
Note the empty argument \redcol\verb${}$\txcol at the end -
without the dots/blobs will not appear!
\vspace{2mm}\\
\begin{minipage}{0.7\linewidth}
\vercol\begin{verbatim}
   ...
   \vertex [dot] (b) at (2,0) {};
   ...
\end{verbatim}\txcol
\end{minipage}
\begin{minipage}{0.25\linewidth}
\tikzsetnextfilename{vertex_dot}
\begin{tikzpicture}
   \begin{feynhand}
   \vertex (a) at (0,0);
   \vertex [dot] (b) at (2,0) {};
   \vertex (c1) at (4,0.5);
   \vertex (c2) at (4,-0.5);
   \propag [plain] (a) to (b);
   \propag [plain] (b) to (c1);
   \propag [plain] (b) to (c2);
   \end{feynhand}
\end{tikzpicture}
\end{minipage}
%
\subsubsection*{Ring dot vertex}
\begin{minipage}{0.7\linewidth}
\vercol\begin{verbatim}
   ...
   \vertex [ringdot] (b) at (2,0) {};
   ...
\end{verbatim}\txcol
\end{minipage}
\begin{minipage}{0.25\linewidth}
\tikzsetnextfilename{vertex_ringdot}
\begin{tikzpicture}
   \begin{feynhand}
   \vertex (a) at (0,0);
   \vertex [ringdot] (b) at (2,0) {};
   \vertex (c1) at (4,0.5);
   \vertex (c2) at (4,-0.5);
   \propag [plain] (a) to (b);
   \propag [plain] (b) to (c1);
   \propag [plain] (b) to (c2);
   \end{feynhand}
\end{tikzpicture}
\end{minipage}
%
\subsubsection*{Square dot vertex}
\begin{minipage}{0.7\linewidth}
\vercol\begin{verbatim}
   ...
   \vertex [squaredot] (b) at (2,0) {};
   ...
\end{verbatim}\txcol
\end{minipage}
\begin{minipage}{0.25\linewidth}
\tikzsetnextfilename{vertex_squaredot}
\begin{tikzpicture}
   \begin{feynhand}
   \vertex (a) at (0,0);
   \vertex [squaredot] (b) at (2,0) {};
   \vertex (c1) at (4,0.5);
   \vertex (c2) at (4,-0.5);
   \propag [plain] (a) to (b);
   \propag [plain] (b) to (c1);
   \propag [plain] (b) to (c2);
   \end{feynhand}
\end{tikzpicture}
\end{minipage}
%
\subsubsection*{Crossed dot vertex}
\begin{minipage}{0.7\linewidth}
\vercol\begin{verbatim}
   ...
   \vertex [crossdot] (b) at (2,0) {};
   ...
\end{verbatim}\txcol
\end{minipage}
\begin{minipage}{0.25\linewidth}
\tikzsetnextfilename{vertex_crossdot}
\begin{tikzpicture}
   \begin{feynhand}
   \vertex (a) at (0,0);
   \vertex [crossdot] (b) at (2,0) {};
   \vertex (c1) at (4,0.5);
   \vertex (c2) at (4,-0.5);
   \propag [plain] (a) to (b);
   \propag [plain] (b) to (c1);
   \propag [plain] (b) to (c2);
   \end{feynhand}
\end{tikzpicture}
\end{minipage}
%
\subsubsection*{Blob vertex}
\begin{minipage}{0.7\linewidth}
\vercol\begin{verbatim}
   ...
   \vertex [blob] (b) at (2,0) {};
   ...
\end{verbatim}\txcol
\end{minipage}
\begin{minipage}{0.25\linewidth}
\tikzsetnextfilename{vertex_blob}
\begin{tikzpicture}
   \begin{feynhand}
   \vertex (a) at (0,0);
   \vertex [blob] (b) at (2,0) {};
   \vertex (c1) at (4,0.5);
   \vertex (c2) at (4,-0.5);
   \propag [plain] (a) to (b);
   \propag [plain] (b) to (c1);
   \propag [plain] (b) to (c2);
   \end{feynhand}
\end{tikzpicture}
\end{minipage}
%
\subsubsection*{Ring blob vertex}
\begin{minipage}{0.7\linewidth}
\vercol\begin{verbatim}
   ...
   \vertex [ringblob] (b) at (2,0) {};
   ...
\end{verbatim}\txcol
\end{minipage}
\begin{minipage}{0.25\linewidth}
\tikzsetnextfilename{vertex_ringblob}
\begin{tikzpicture}
   \begin{feynhand}
   \vertex (a) at (0,0);
   \vertex [ringblob] (b) at (2,0) {};
   \vertex (c1) at (4,0.5);
   \vertex (c2) at (4,-0.5);
   \propag [plain] (a) to (b);
   \propag [plain] (b) to (c1);
   \propag [plain] (b) to (c2);
   \end{feynhand}
\end{tikzpicture}
\end{minipage}
%
\subsubsection*{Gray blob vertex}
\begin{minipage}{0.7\linewidth}
\vercol\begin{verbatim}
   ...
   \vertex [grayblob] (b) at (2,0) {};
   ...
\end{verbatim}\txcol
\end{minipage}
\begin{minipage}{0.25\linewidth}
\tikzsetnextfilename{vertex_grayblob}
\begin{tikzpicture}
   \begin{feynhand}
   \vertex (a) at (0,0);
   \vertex [grayblob] (b) at (2,0) {};
   \vertex (c1) at (4,0.5);
   \vertex (c2) at (4,-0.5);
   \propag [plain] (a) to (b);
   \propag [plain] (b) to (c1);
   \propag [plain] (b) to (c2);
   \end{feynhand}
\end{tikzpicture}
\end{minipage}
%
\subsubsection*{NorthWest blob vertex}
\begin{minipage}{0.7\linewidth}
\vercol\begin{verbatim}
   ...
   \vertex [NWblob] (b) at (2,0) {};
   ...
\end{verbatim}\txcol
\end{minipage}
\begin{minipage}{0.25\linewidth}
\tikzsetnextfilename{vertex_NWblob}
\begin{tikzpicture}
   \begin{feynhand}
   \vertex (a) at (0,0);
   \vertex [NWblob] (b) at (2,0) {};
   \vertex (c1) at (4,0.5);
   \vertex (c2) at (4,-0.5);
   \propag [plain] (a) to (b);
   \propag [plain] (b) to (c1);
   \propag [plain] (b) to (c2);
   \end{feynhand}
\end{tikzpicture}
\end{minipage}
%
\subsubsection*{NorthEast blob vertex}
\begin{minipage}{0.7\linewidth}
\vercol\begin{verbatim}
   ...
   \vertex [NEblob] (b) at (2,0) {};
   ...
\end{verbatim}\txcol
\end{minipage}
\begin{minipage}{0.25\linewidth}
\tikzsetnextfilename{vertex_NEblob}
\begin{tikzpicture}
   \begin{feynhand}
   \vertex (a) at (0,0);
   \vertex [NEblob] (b) at (2,0) {};
   \vertex (c1) at (4,0.5);
   \vertex (c2) at (4,-0.5);
   \propag [plain] (a) to (b);
   \propag [plain] (b) to (c1);
   \propag [plain] (b) to (c2);
   \end{feynhand}
\end{tikzpicture}
\end{minipage}
%
\subsubsection{Relative vertex placement}
\label{sec:_feynmandiagrams_vertices_relative}
\noindent
Above we have always placed the vertices using absolute coordinate values.
Relative placement is done using the keys
\blucol\verb$above, below, left, right$\txcol.
When combined, the vertical key must come first, as in
\blucol\verb$above right$\txcol.\\

\noindent
\begin{minipage}{0.75\linewidth}
\blucol\begin{verbatim}
   \vertex [dot] (a1) {};
   \vertex [ringdot] (a2) [above right = of a1] {};
   \propag [plain] (a1) to (a2);
\end{verbatim}\txcol
\end{minipage}
\begin{minipage}{0.2\linewidth}
\tikzsetnextfilename{vertex_relative_1}
\begin{tikzpicture}
   \begin{feynhand}
   \vertex [dot] (a1) {};
   \vertex [ringdot] (a2) [above right = of a1] {};
   \propag [plain] (a1) to (a2);
   \end{feynhand}
\end{tikzpicture}
\end{minipage}
\\

\noindent
The node distance used by these keys can be set
by calling \blucol\verb$\begin{tikzpicture}[node distance = 2cm]$\txcol
or \blucol\verb$\begin{feynhand}[node distance = 2cm]$\txcol.
Moreover, we can specify the absolute distances relative to a vertex:\\

\noindent
\begin{minipage}{0.75\linewidth}
\blucol\begin{verbatim}
   \vertex [dot] (a1) {};
   \vertex [dot] (a2) [right = 1.5cm of a1] {};
   \propag [gluon] (a1) to (a2);
\end{verbatim}\txcol
\end{minipage}
\begin{minipage}{0.2\linewidth}
\tikzsetnextfilename{vertex_relative_2}
\begin{tikzpicture}
   \begin{feynhand}
   \vertex [dot] (a1) {};
   \vertex [dot] (a2) [right = 1.5cm of a1] {};
   \propag [gluon] (a1) to (a2);
   \end{feynhand}
\end{tikzpicture}
\end{minipage}
\\
\begin{minipage}{0.75\linewidth}
\blucol\begin{verbatim}
   \vertex [ringdot] (a1) {};
   \vertex [ringdot] (a2) [above right = 0.5cm and 2cm of a1] {};
   \propag [photon] (a1) to (a2);
\end{verbatim}\txcol
\end{minipage}
\begin{minipage}{0.2\linewidth}
\tikzsetnextfilename{vertex_relative_3}
\begin{tikzpicture}
   \begin{feynhand}
   \vertex [ringdot] (a1) {};
   \vertex [ringdot] (a2) [above right = 0.5cm and 2cm of a1] {};
   \propag [photon] (a1) to (a2);
   \end{feynhand}
\end{tikzpicture}
\end{minipage}

%
%
\subsection{Propagators}
\label{sec:_feynmandiagrams_propagators}
\noindent
You can draw propagators between vertices with FeynHand's \vercol\verb$\propag$\txcol command,
whose long version \vercol\verb$\propagator$\txcol also works.
Both are just TikZ' \vercol\verb$\draw$\txcol command, 
and thus can access all styles defined in TikZ-Feynman.
The minimal example \vercol\verb$\propag[fer] (a) to (b);$\txcol
draws a fermion propagator from vertex (a) and (b).
Below we list all propagator styles and draw examples.
FeynHand uses intuitive 3/6-letter versions, but also keeps TikZ-Feynman's longer ones:
\setlength{\extrarowheight}{3pt}
\begin{table}[H]
	\centering
	\begin{tabular}{r>{\qquad} r}
		\hline
		Short Version & Long Version
			\\
		\hhline{==}
		\verb$\propag$ & \verb$\propagator$\\
		\hline
		\verb$fer$ & \verb$fermion$\\
		\verb$antfer$ & \verb$anti fermion$\\
		\hline
		\verb$pho$ & \verb$photon$\\
		\hline
		\verb$bos$ & \verb$boson$\\
		\verb$chabos$ & \verb$charged boson$\\
		\verb$antbos$ & \verb$anti charged boson$\\
		\hline
		\verb$glu$ & \verb$gluon$\\
		\hline
	\end{tabular}
	\qquad\qquad\qquad
	\begin{tabular}{r>{\qquad} r}
		\hline
		Short Version & Long Version
			\\
		\hhline{==}
		\verb$sca$ & \verb$scalar$\\
		\verb$chasca$ & \verb$charged scalar$\\
		\verb$antsca$ & \verb$anti charged scalar$\\
		\hline
		\verb$gho$ & \verb$ghost$\\
		\verb$chagho$ & \verb$charged ghost$\\
		\verb$antgho$ & \verb$anti charged ghost$\\
		\hline
		\verb$maj$ & \verb$majorana$\\
		\verb$antmaj$ & \verb$anti majorana$\\
		\hline
	\end{tabular}
\end{table}
%
%
\subsubsection*{Fermions}
\begin{minipage}{0.7\linewidth}
\vercol\begin{verbatim}
   \vertex (a) at (0,0); \vertex (b) at (2,0);
   \propag[fer] (a) to (b);
\end{verbatim}\txcol
\end{minipage}
\begin{minipage}{0.25\linewidth}
\tikzsetnextfilename{propag_fermion}
\begin{tikzpicture}
   \begin{feynhand}
   \vertex (a) at (0,0); \vertex (b) at (2,0);
   \propag[fer] (a) to (b);
   \end{feynhand}
\end{tikzpicture}
\end{minipage}
\\
\begin{minipage}{0.7\linewidth}
\vercol\begin{verbatim}
   \vertex (a) at (0,0); \vertex (b) at (2,0);
   \propagator[antfer] (a) to (b);
\end{verbatim}\txcol
\end{minipage}
\begin{minipage}{0.25\linewidth}
\tikzsetnextfilename{propag_antfer}
\begin{tikzpicture}
   \begin{feynhand}
   \vertex (a) at (0,0); \vertex (b) at (2,0);
   \propagator[antfer] (a) to (b);
   \end{feynhand}
\end{tikzpicture}
\end{minipage}
%
\subsubsection*{Gluons}
\begin{minipage}{0.7\linewidth}
\vercol\begin{verbatim}
   \vertex (a) at (0,0); \vertex (b) at (2,0);
   \propag[glu] (a) to (b);
\end{verbatim}\txcol
\end{minipage}
\begin{minipage}{0.25\linewidth}
\tikzsetnextfilename{propag_gluon}
\begin{tikzpicture}
   \begin{feynhand}
   \vertex (a) at (0,0); \vertex (b) at (2,0);
   \propag[glu] (a) to (b);
   \end{feynhand}
\end{tikzpicture}
\end{minipage}
\\
\begin{minipage}{0.7\linewidth}
\vercol\begin{verbatim}
   \vertex (a) at (0,0); \vertex (b) at (2,1);
   \propag[glu] (a) to [out=0, in=180] (b);
\end{verbatim}\txcol
\end{minipage}
\begin{minipage}{0.25\linewidth}
\tikzsetnextfilename{propag_gluon_out_in}
\begin{tikzpicture}
   \begin{feynhand}
   \vertex (a) at (0,0); \vertex (b) at (2,1);
   \propag[gluon] (a) to [out=0, in=180] (b);
   \end{feynhand}
\end{tikzpicture}
\end{minipage}
\\
%
\subsubsection*{Bosons (photons are just bosons)}
\begin{minipage}{0.7\linewidth}
\vercol\begin{verbatim}
   \vertex (a) at (0,0); \vertex (b) at (2,0);
   \propag[bos] (a) to (b);
\end{verbatim}\txcol
\end{minipage}
\begin{minipage}{0.25\linewidth}
\tikzsetnextfilename{propag_boson}
\begin{tikzpicture}
   \begin{feynhand}
   \vertex (a) at (0,0); \vertex (b) at (2,0);
   \propag[bos] (a) to (b);
   \end{feynhand}
\end{tikzpicture}
\end{minipage}
\\
\begin{minipage}{0.7\linewidth}
\vercol\begin{verbatim}
   \vertex (a) at (0,0); \vertex (b) at (2,0);
   \propag[pho] (a) to (b);
\end{verbatim}\txcol
\end{minipage}
\begin{minipage}{0.25\linewidth}
\tikzsetnextfilename{propag_photon}
\begin{tikzpicture}
   \begin{feynhand}
   \vertex (a) at (0,0); \vertex (b) at (2,0);
   \propag[pho] (a) to (b);
   \end{feynhand}
\end{tikzpicture}
\end{minipage}
\\
\begin{minipage}{0.7\linewidth}
\vercol\begin{verbatim}
   \vertex (a) at (0,0); \vertex (b) at (2,-1);
   \propag[bos] (a) to [out=0, in=90] (b);
\end{verbatim}\txcol
\end{minipage}
\begin{minipage}{0.25\linewidth}
\tikzsetnextfilename{propag_boson_out_in}
\begin{tikzpicture}
   \begin{feynhand}
   \vertex (a) at (0,0); \vertex (b) at (2,-1);
   \propag[bos] (a) to [out=0, in=90] (b);
   \end{feynhand}
\end{tikzpicture}
\end{minipage}
\\
%
\subsubsection*{Charged bosons}
\begin{minipage}{0.7\linewidth}
\vercol\begin{verbatim}
   \vertex (a) at (0,0); \vertex (b) at (2,0);
   \propag[chabos] (a) to (b);
\end{verbatim}\txcol
\end{minipage}
\begin{minipage}{0.25\linewidth}
\tikzsetnextfilename{propag_chabos}
\begin{tikzpicture}
   \begin{feynhand}
   \vertex (a) at (0,0); \vertex (b) at (2,0);
   \propag[chabos] (a) to (b);
   \end{feynhand}
\end{tikzpicture}
\end{minipage}
\\
\begin{minipage}{0.7\linewidth}
\vercol\begin{verbatim}
   \vertex (a) at (0,0); \vertex (b) at (2,0);
   \propag[antbos] (a) to (b);
\end{verbatim}\txcol
\end{minipage}
\begin{minipage}{0.25\linewidth}
\tikzsetnextfilename{propag_antbos}
\begin{tikzpicture}
   \begin{feynhand}
   \vertex (a) at (0,0); \vertex (b) at (2,0);
   \propag[antbos] (a) to (b);
   \end{feynhand}
\end{tikzpicture}
\end{minipage}
\\
%
\newpage
\subsubsection*{Scalars}
\begin{minipage}{0.7\linewidth}
\vercol\begin{verbatim}
   \vertex (a) at (0,0); \vertex (b) at (2,0);
   \propag[sca] (a) to (b);
\end{verbatim}\txcol
\end{minipage}
\begin{minipage}{0.25\linewidth}
\tikzsetnextfilename{propag_scalar}
\begin{tikzpicture}
   \begin{feynhand}
   \vertex (a) at (0,0); \vertex (b) at (2,0);
   \propag[sca] (a) to (b);
   \end{feynhand}
\end{tikzpicture}
\end{minipage}
\\
\begin{minipage}{0.7\linewidth}
\vercol\begin{verbatim}
   \vertex (a) at (0,0); \vertex (b) at (2,1);
   \propag[sca] (a) to [out=0, in=180] (b);
\end{verbatim}\txcol
\end{minipage}
\begin{minipage}{0.25\linewidth}
\tikzsetnextfilename{propag_scalar_out_in}
\begin{tikzpicture}
   \begin{feynhand}
   \vertex (a) at (0,0); \vertex (b) at (2,1);
   \propag[sca] (a) to [out=0, in=180] (b);
   \end{feynhand}
\end{tikzpicture}
\end{minipage}
\\
%
\vspace{-6mm}
\subsubsection*{Charged scalars}
\begin{minipage}{0.7\linewidth}
\vercol\begin{verbatim}
   \vertex (a) at (0,0); \vertex (b) at (2,0);
   \propag[chasca] (a) to (b);
\end{verbatim}\txcol
\end{minipage}
\begin{minipage}{0.25\linewidth}
\tikzsetnextfilename{propag_chasca}
\begin{tikzpicture}
   \begin{feynhand}
   \vertex (a) at (0,0); \vertex (b) at (2,0);
   \propag[chasca] (a) to (b);
   \end{feynhand}
\end{tikzpicture}
\end{minipage}
\\
\begin{minipage}{0.7\linewidth}
\vercol\begin{verbatim}
   \vertex (a) at (0,0); \vertex (b) at (2,0);
   \propag[antsca] (a) to (b);
\end{verbatim}\txcol
\end{minipage}
\begin{minipage}{0.25\linewidth}
\tikzsetnextfilename{propag_antsca}
\begin{tikzpicture}
   \begin{feynhand}
   \vertex (a) at (0,0); \vertex (b) at (2,0);
   \propag[antsca] (a) to (b);
   \end{feynhand}
\end{tikzpicture}
\end{minipage}
\\
%
\subsubsection*{Ghosts}
\begin{minipage}{0.7\linewidth}
\vercol\begin{verbatim}
   \vertex (a) at (0,0); \vertex (b) at (2,0);
   \propag[gho] (a) to (b);
\end{verbatim}\txcol
\end{minipage}
\begin{minipage}{0.25\linewidth}
\tikzsetnextfilename{propag_ghost}
\begin{tikzpicture}
   \begin{feynhand}
   \vertex (a) at (0,0); \vertex (b) at (2,0);
   \propag[gho] (a) to (b);
   \end{feynhand}
\end{tikzpicture}
\end{minipage}
\\
\begin{minipage}{0.7\linewidth}
\vercol\begin{verbatim}
   \vertex (a) at (0,0); \vertex (b) at (2,1);
   \propag[gho] (a) to [out=0, in=180] (b);
\end{verbatim}\txcol
\end{minipage}
\begin{minipage}{0.25\linewidth}
\tikzsetnextfilename{propag_ghost_out_in}
\begin{tikzpicture}
   \begin{feynhand}
   \vertex (a) at (0,0); \vertex (b) at (2,1);
   \propag[gho] (a) to [out=0, in=180] (b);
   \end{feynhand}
\end{tikzpicture}
\end{minipage}
%
\subsubsection*{Charged ghosts}
\begin{minipage}{0.7\linewidth}
\vercol\begin{verbatim}
   \vertex (a) at (0,0); \vertex (b) at (2,0);
   \propag[chagho] (a) to (b);
\end{verbatim}\txcol
\end{minipage}
\begin{minipage}{0.25\linewidth}
\tikzsetnextfilename{propag_chaghost}
\begin{tikzpicture}
   \begin{feynhand}
   \vertex (a) at (0,0); \vertex (b) at (2,0);
   \propag[chagho] (a) to (b);
   \end{feynhand}
\end{tikzpicture}
\end{minipage}
\\
\begin{minipage}{0.7\linewidth}
\vercol\begin{verbatim}
   \vertex (a) at (0,0); \vertex (b) at (2,0);
   \propag[antgho] (a) to (b);
\end{verbatim}\txcol
\end{minipage}
\begin{minipage}{0.25\linewidth}
\tikzsetnextfilename{propag_antghost}
\begin{tikzpicture}
   \begin{feynhand}
   \vertex (a) at (0,0); \vertex (b) at (2,0);
   \propag[antgho] (a) to (b);
   \end{feynhand}
\end{tikzpicture}
\end{minipage}
\\
%
\subsubsection*{Majorana}
\begin{minipage}{0.7\linewidth}
\vercol\begin{verbatim}
   \vertex (a) at (0,0); \vertex (b) at (2,0);
   \propag[maj] (a) to (b);
\end{verbatim}\txcol
\end{minipage}
\begin{minipage}{0.25\linewidth}
\tikzsetnextfilename{propag_major}
\begin{tikzpicture}
   \begin{feynhand}
   \vertex (a) at (0,0); \vertex (b) at (2,0);
   \propag[maj] (a) to (b);
   \end{feynhand}
\end{tikzpicture}
\end{minipage}
\\
\begin{minipage}{0.7\linewidth}
\vercol\begin{verbatim}
   \vertex (a) at (0,0); \vertex (b) at (2,0);
   \propag[antmaj] (a) to (b);
\end{verbatim}\txcol
\end{minipage}
\begin{minipage}{0.25\linewidth}
\tikzsetnextfilename{propag_antmaj}
\begin{tikzpicture}
   \begin{feynhand}
   \vertex (a) at (0,0); \vertex (b) at (2,0);
   \propag[antmaj] (a) to (b);
   \end{feynhand}
\end{tikzpicture}
\end{minipage}
\\
%
\subsubsection*{Linked propagators}
\noindent
We can also draw "chains" of propagators from one vertex to the next to the next etc.~using
TikZ' \blucol\verb$\graph$\txcol command. 
Then, all style options are attached to the connector \blucol\verb$--$\txcol 
as in the following example. 
\\

\begin{minipage}{0.75\linewidth}
\blucol\begin{verbatim}
   \graph {(a0) --[fer,green!50!black] 
           (a1) --[glu,red] 
           (a2) --[chabos,blue] (a3)};
           
   \graph {(b0) --[fer, red, in=180, out=270, insertion=0.8] 
           (b1) --[glu, edge label =$k$] 
           (b2) --[chabos, blue, mom={[arrow style=blue] $q$}] (b3)};
\end{verbatim}\txcol
\end{minipage}
\begin{minipage}{0.23\linewidth}
\tikzsetnextfilename{propag_chain}
\begin{tikzpicture}
   \begin{feynhand}
   \vertex (a0) at (0,2); \vertex (a1) at (1,2); 
   \vertex (a2) at (2,2); \vertex (a3) at (3,2);
   \vertex (b0) at (0,1); \vertex (b1) at (1,0); 
   \vertex (b2) at (2,0); \vertex (b3) at (3,1);
   \graph {(a0) --[fer, green!50!black] (a1) --[glu, red] (a2) --[chabos, blue] (a3)};
   \graph {(b0) --[fer, red, in=180, out=270, insertion=0.8] 
   (b1) --[glu, edge label =$k$] (b2) --[chabos, blue, mom={[arrow style=blue] $q$}] (b3)};
   \end{feynhand}
\end{tikzpicture}
\end{minipage}
%
\newpage\subsubsection{Keys: In, Out, Looseness, Left/Right, Half/Quarter}
\label{sec:_feynmandiagrams_keysinoutetc}
\noindent
The keys \vercol\verb$in, out$\txcol specify the angle
at which the line leaves and enters the vertices.
The angle is with respect to the coordinate system:
0 degrees is right, 90 up, 180 left, 270 down.
\vspace{2mm}\\
\begin{minipage}{0.7\linewidth}
\vercol\begin{verbatim}
   \vertex (a) at (0,0); \vertex (b) at (2,0);
   \propag[fer] (a) to [out=90, in=90] (b);
\end{verbatim}\txcol
\end{minipage}
\begin{minipage}{0.25\linewidth}
\tikzsetnextfilename{propag_keys_out_in}
\begin{tikzpicture}
   \begin{feynhand}
   \vertex (a) at (0,0); \vertex (b) at (2,0);
   \propag[fer] (a) to [out=90, in=90] (b);
   \end{feynhand}
\end{tikzpicture}
\end{minipage}
\\
The key \vercol\verb$looseness$\txcol specifies
how loose (distant) the arc is bending away
from the straight line connecting the vertices.
Its default value is 1.
\vspace{2mm}\\
\begin{minipage}{0.7\linewidth}
\vercol\begin{verbatim}
   \vertex (a) at (0,0); \vertex (b) at (2,0);
   \vertex (c) at (3,0); \vertex (d) at (5,0);
   \propag[fer] (a) to [in=90, out=90] (b);
   \propag[fer] (c) to [in=90, out=90, looseness=1.5] (d);
\end{verbatim}\txcol
\end{minipage}
\begin{minipage}{0.25\linewidth}
\tikzsetnextfilename{propag_keys_looseness}
\begin{tikzpicture}
   \begin{feynhand}
   \vertex (a) at (0,0); \vertex (b) at (2,0);
   \vertex (c) at (3,0); \vertex (d) at (5,0);
   \propag[fer] (a) to [in=90, out=90] (b);
   \propag[fer] (c) to [in=90, out=90, looseness=1.5] (d);
   \end{feynhand}
\end{tikzpicture}
\end{minipage}
\\
The keys \vercol\verb$half left, half right, quarter left, quarter right$\txcol
specify that the line is a half/quarter circle bending to the left/right
with respect to the line's direction.
\vspace{2mm}\\
\begin{minipage}{0.7\linewidth}
\vercol\begin{verbatim}
   \vertex (a) at (0,0); \vertex (b) at (2,0);
   \vertex (c) at (3,1); \vertex (d) at (5,1);
   \propag[fer] (a) to [half left] (b);
   \propag[fer] (c) to [half right, looseness=1.5] (d);
\end{verbatim}\txcol
\end{minipage}
\begin{minipage}{0.25\linewidth}
\tikzsetnextfilename{propag_keys_half}
\begin{tikzpicture}
   \begin{feynhand}
   \vertex (a) at (0,0); \vertex (b) at (2,0);
   \vertex (c) at (3,1); \vertex (d) at (5,1);
   \propag[fer] (a) to [half left] (b);
   \propag[fer] (c) to [half right, looseness=1.5] (d);
   \end{feynhand}
\end{tikzpicture}
\end{minipage}
\\
\begin{minipage}{0.7\linewidth}
\vercol\begin{verbatim}
   \vertex (a) at (0,0); \vertex (b) at (2,0);
   \vertex (c) at (3,1); \vertex (d) at (4,0);
   \propag[fer] (a) to [quarter left] (b);
   \propag[fer] (c) to [quarter right, looseness=1.5] (d);
\end{verbatim}\txcol
\end{minipage}
\begin{minipage}{0.25\linewidth}
\tikzsetnextfilename{propag_keys_quarter}
\begin{tikzpicture}
   \begin{feynhand}
   \vertex (a) at (0,0); \vertex (b) at (2,0);
   \vertex (c) at (3,1); \vertex (d) at (4,0);
   \propag[fer] (a) to [quarter left] (b);
   \propag[fer] (c) to [quarter right, looseness=1.5] (d);
   \end{feynhand}
\end{tikzpicture}
\end{minipage}
%
\vspace{-6mm}
\subsubsection{Propagator labels and momentum arrows}
\label{sec:_feynmandiagrams_proplabelsmomentum}
\noindent
In order to include a label to a propagator line,
we can use TikZ's \blucol\verb!edge label!\txcol key,
see Section~14.13 in the TikZ/PGF Manual \cite{tikz}.
\blucol\verb!edge label!\txcol sets a propagator label 
to the left of the propagator line, and the primed version
\blucol\verb!edge label'!\txcol sets one to its right,
see the following example.
Left and right are relative with respect to the line's direction.
\vspace{2mm}\\
\begin{minipage}{0.83\linewidth}
\blucol\begin{verbatim}
 \vertex (a1) at (0,0); \vertex (a2) at (1,1.5);
 \vertex (b1) at (1.5,1.5); \vertex (b2) at (2.5,0);
 \propag [fer, Orange]  (a1) to [edge label = $k$] (a2);
 \propag [fer, Blue] (b1) to [edge label'= $p'$] (b2);
\end{verbatim}\txcol
\end{minipage}
\begin{minipage}{0.16\linewidth}
\tikzsetnextfilename{propag_label}
\begin{tikzpicture}
   \begin{feynhand}
   \vertex (a1) at (0,0); \vertex (a2) at (1,1.5);
   \vertex (b1) at (1.5,1.5); \vertex (b2) at (2.5,0);
   \propag [fer, RedOrange]  (a1) to [edge label = $k$] (a2);
   \propag [fer, Blue] (b1) to [edge label'= $p'$] (b2);
   \end{feynhand}
\end{tikzpicture}
\end{minipage}
\\
In order to add arrows indicating the momentum's direction to a propagator line,
we can use the key \blucol\verb!momentum!\txcol.
The key \blucol\verb!reversed momentum!\txcol draws the arrow in the reverse direction,
and a prime changes the side on which the arrow appears, as above.
FeynHand also offers the 3/6-letter shorthands \blucol\verb!mom!\txcol 
and \blucol\verb!revmom!\txcol.
The momentum arrows \emph{do not} inherit the drawing style of the propagator line,
and hence their style can be specified by an extra argument.
\vspace{2mm}\\
\begin{minipage}{0.83\linewidth}
\blucol\begin{verbatim}
 \vertex (a1) at (0,0); \vertex (a2) at (1,1.5);
 \vertex (b1) at (1.5,1.5); \vertex (b2) at (2.5,0);
 \propag [fer, Orange, mom={[arrow style=Orange] $k$}] (a1) to (a2);
 \propag [fer, Blue, revmom'={[arrow style=Blue] $p'$}] (b1) to (b2);
\end{verbatim}\txcol
\end{minipage}
\begin{minipage}{0.16\linewidth}
\tikzsetnextfilename{propag_arrows}
\begin{tikzpicture}
   \begin{feynhand}
   \vertex (a1) at (0,0); \vertex (a2) at (1,1.5);
   \vertex (b1) at (1.5,1.5); \vertex (b2) at (2.5,0);
   \propag [fer, RedOrange, mom={[arrow style=RedOrange] $k$}] (a1) to (a2);
   \propag [fer, Blue, revmom'={[arrow style=Blue] $p'$}] (b1) to (b2);
   \end{feynhand}
\end{tikzpicture}
\end{minipage}
\\
Note that \blucol\verb!edge label!\txcol 
is an option of the \blucol\verb!to!\txcol key,
whereas \blucol\verb!momentum!\txcol is an option 
of the \blucol\verb!\propag!\txcol command.
%
\vspace{-3mm}
\subsubsection{Insertions}
\label{sec:_feynmandiagrams_propagators_insertions}
\noindent
Insertions can be placed on propagator lines as in the examples below.
The values indicate the fraction of the propagator line's length
at which the insertion marks are drawn.
\\
\begin{minipage}{0.83\linewidth}
\vercol\begin{verbatim}
  \vertex [dot] (a1) at (0,0) {};   \vertex [dot] (a2) at (2,0) {};
  \propag [chabos, red, insertion=0.25, insertion=0.75] (a1) to (a2);
\end{verbatim}\txcol
\end{minipage}
\begin{minipage}{0.16\linewidth}
\tikzsetnextfilename{propag_insertion_1}
\begin{tikzpicture}
   \begin{feynhand}
  \vertex [dot] (a1) at (0,0) {};
  \vertex [dot] (a2) at (2,0) {};
  \propag [chabos, red, insertion=0.25, insertion=0.75] (a1) to (a2);
  \end{feynhand}
\end{tikzpicture}
\end{minipage}
\\
\begin{minipage}{0.83\linewidth}
\vercol\begin{verbatim}
  \vertex [ringdot] (a1) at (0,0) {};   \vertex [ringdot] (a2) at (2,0) {};
  \propag [chabos, red, insertion={[size=6pt,style=Green]0.25}] (a1) to (a2);
\end{verbatim}\txcol
\end{minipage}
\begin{minipage}{0.16\linewidth}
\tikzsetnextfilename{propag_insertion_2}
\begin{tikzpicture}
   \begin{feynhand}
  \vertex [ringdot] (a1) at (0,0) {};
  \vertex [ringdot] (a2) at (2,0) {};
  \propag [chabos, red, insertion={[size=6pt,style=Green]0.25}] (a1) to (a2);
   \end{feynhand}
\end{tikzpicture}
\end{minipage}

%
\subsubsection{Crossing propagators (one on top of the other)}
\label{sec:_feynmandiagrams_propagators_top}
\noindent
Crossings often indicate that the propagators merely cross on paper
without any physical interaction being implied.
First, draw the propagator line to become crossed/interrupted,
then draw the crossing propagator on top of it by specifying 
the additional style key \blucol\verb$top$\txcol.
For more intricate networks of crossing propagators, you can place auxiliary vertices 
to break the propagators into parts if necessary, 
and then place the respective parts on top of each other.

If due to the crossings you wish to move the arrows of charged particles,
specify \blucol\verb$with arrow=VALUE$\txcol,
respectively \blucol\verb$with reversed arrow=VALUE$\txcol for antiparticles.
\blucol\verb$VALUE$\txcol must be a number between 0 and 1,
and indicates the fraction of propagator length at which TikZ inserts the arrow.
The default value of \blucol\verb$VALUE$\txcol is 0.5.
\vspace{2mm}\\
\begin{minipage}{0.8\linewidth}
\blucol\begin{verbatim}
   \vertex (a) at (0,0); \vertex (b) at (2,1);
   \vertex (c) at (0,1); \vertex (d) at (2,0);
   \propag [chabos, Orange, with arrow=0.25] (c) to (d);
   \propag [fer, Blue, top] (a) to (b);
\end{verbatim}\txcol
\end{minipage}
\begin{minipage}{0.15\linewidth}
\tikzsetnextfilename{propag_gap_up}
\begin{tikzpicture}
   \begin{feynhand}
   \vertex (a) at (0,0); \vertex (b) at (2,1);
   \vertex (c) at (0,1); \vertex (d) at (2,0);
   \propag [chabos, Orange, with arrow=0.25] (c) to (d);
   \propag [fer, Blue, top] (a) to (b);
   \end{feynhand}
\end{tikzpicture}
\end{minipage}
\\
\begin{minipage}{0.8\linewidth}
\blucol\begin{verbatim}
   ...
   \propag[gluon, blue] (a) to [out=0, in=180] (b);
   \propag[gluon, top] (c) to [out=0, in=180] (d);
\end{verbatim}\txcol
\end{minipage}
\begin{minipage}{0.15\linewidth}
\tikzsetnextfilename{propag_gluon_gap}
\begin{tikzpicture}
   \begin{feynhand}
   \vertex (a) at (0,0); \vertex (b) at (2,1);
   \vertex (c) at (0,1); \vertex (d) at (2,0);
   \propag[gluon, blue] (a) to [out=0, in=180] (b);
   \propag[gluon, top] (c) to [out=0, in=180] (d);
   \end{feynhand}
\end{tikzpicture}
\end{minipage}
%
%
%
\subsection{Customizing styles}
\label{sec:_feynmandiagrams_customizing}
FeynHand provides the following shortcuts for quickly modifying styles.
Note that any change of these lengths at some point in the code
affects all diagrams that come after this point.
Also recall that if you externalize graphics (see Sec.~\ref{Apx_tikzbasics_externalize}),
then TikZ refreshes pictures \emph{only} when their source code is changed.
Hence to see the effect of such length changes, 
delete the external PDF-files or use \vercol\verb$\tikzset{external/force remake}$\txcol.
\begin{itemize}
	\item
	\vercol\verb$\setlength{\feynhanddotsize}{2mm}$\txcol
	sets the diameter of the dots to 2mm (default: 1.5mm).
	\item
	\vercol\verb$\setlength{\feynhandblobsize}{10mm}$\txcol
	sets the diameter of the blobs to 10mm (default: 7.5mm).
	\item
	\vercol\verb$\setlength{\feynhandlinesize}{2pt}$\txcol
	sets the thickness of the propagator lines to 2pt (default: 0.5pt).\\
	Also the lines of dots and blobs use this thickness.\\
	The momentum arrows' line thickness scales as 0.64 times this value.
	Also the amplitudes\\
	and section lengths of the boson and photon waves
	and of the gluon curls scale with this value.
	\item
	\vercol\verb$\setlength{\feynhandarrowsize}{8pt}$\txcol
	sets the size of the arrows to 8pt (default: 6pt)\\
	on all charged propagator lines
	\vercol\verb$(fer, antfer, chabos, antbos, chasca, antsca, chagho, antgho)$\txcol\\
	and majorana lines \vercol\verb$(maj, antmaj).$\txcol
	The momentum arrows' tip size scales as 0.8 times this value.
	\item
	\vercol\verb$\setlength{\feynhandtopsep}{8pt}$\txcol
	sets the width of the crossing gap for propagators on top of each other.
	The default value is \vercol\verb$18\feynhandlinesize$\txcol.
	\item
	\vercol\verb$\renewcommand{\feynhandtopsepcolor}{yellow}$\txcol
	sets the color of the crossing gap (default is white).	
\end{itemize}
\noindent
\begin{minipage}{0.6\linewidth}
Example diagram drawn with default values:
\end{minipage}
\begin{minipage}{0.38\linewidth}
\tikzsetnextfilename{tikz_customize_default}
\begin{tikzpicture}
\begin{feynhand}
  \vertex [ringdot] (a) at (0,0) {};
  \vertex [NWblob] (b) at (2,0) {};
  \vertex [crossdot] (c1) at (4,0.5) {};
  \vertex [dot] (c2) at (5,-0.5) {}; 
  \vertex [dot] (d) at (3,-0.5) {}; 
  \propag [chasca, Green] (a) to (b);
  \propag [gluon, Blue] (b) to (c2);
  \propag [fer, top] (d) to (c1);
  \propag [chabos, RedOrange, momentum={[arrow style=RedOrange] $k$}] (b) to (c1);
\end{feynhand}
\end{tikzpicture}
\end{minipage}
\\
\noindent
\begin{minipage}{0.6\linewidth}
Same diagram (same TikZ/FeynHand code) drawn after setting:
\vercol\begin{verbatim}
\setlength{\feynhanddotsize}{2mm}
\setlength{\feynhandblobsize}{10mm}
\setlength{\feynhandlinesize}{1pt}
\setlength{\feynhandarrowsize}{9pt}
\setlength{\feynhandtopsep}{3mm}
\renewcommand{\feynhandtopsepcolor}{yellow}
\end{verbatim}\txcol
\end{minipage}
\setlength{\feynhanddotsize}{2mm}
\setlength{\feynhandblobsize}{10mm}
\setlength{\feynhandlinesize}{1pt}
\setlength{\feynhandarrowsize}{9pt}
\setlength{\feynhandtopsep}{3mm}
\renewcommand{\feynhandtopsepcolor}{yellow}
\begin{minipage}{0.38\linewidth}
\tikzsetnextfilename{tikz_customize_big}
\begin{tikzpicture}
\begin{feynhand}
  \vertex [ringdot] (a) at (0,0) {};
  \vertex [NWblob] (b) at (2,0) {};
  \vertex [crossdot] (c1) at (4,0.5) {};
  \vertex [dot] (c2) at (5,-0.5) {}; 
  \vertex [dot] (d) at (3,-0.5) {}; 
  \propag [chasca, Green] (a) to (b);
  \propag [gluon, Blue] (b) to (c2);
  \propag [fer, top] (d) to (c1);
  \propag [chabos, RedOrange, momentum={[arrow style=RedOrange] $k$}] (b) to (c1);
\end{feynhand}
\end{tikzpicture}
\end{minipage}
%
\setlength{\feynhanddotsize}{1.5mm}
\setlength{\feynhandblobsize}{7.5mm}
\setlength{\feynhandlinesize}{0.5pt}
\setlength{\feynhandarrowsize}{6pt}
\setlength{\feynhandtopsep}{18\feynhandlinesize}
\renewcommand{\feynhandtopsepcolor}{white}
The color of a vertex/propagator style can be set for all subsequent diagrams
in the current local Tex group (e.g. document or equation, figure, minipage environment) 
by the command \vercol\verb$\tikzfeynhandset$\txcol with the key \vercol\verb$every$\txcol:
\vercol\begin{verbatim}
\tikzfeynhandset{every particle={/tikz/color=blue}, every dot={/tikz/color=red},}
\tikzfeynhandset{every fermion={/tikz/color=green},}
\end{verbatim}\txcol

%
%
\appendix
\section{TikZ Basics}
\label{Apx_tikzbasics}
This appendix summarizes a few things that can be done with TikZ,
just enough for getting started quickly.\\
You can discover much more, and I recommend
becoming familiar with the TikZ/PGF Manual \cite{tikz} bit by bit.

\subsection{Environment}
\label{Apx_tikzbasics_environment}
\noindent
All TikZ commands are used within the 
\brocol\verb$tikzpicture$\txcol environment.
All TikZ commands that we mention are understood 
as appearing in this environment.
When we want to use captions and labels, we enclose it in the 
\brocol\verb$figure$\txcol environment, but it can be called 
anywhere in the text or math of a Tex document:\\

\noindent
\begin{minipage}{0.6\linewidth}
\brocol\begin{verbatim}
\begin{figure}
   \centering
%\tikzset{external/force remake}
   \begin{tikzpicture}
		
      ... TikZ commands ...

   \end{tikzpicture}
   \caption{A diagram drawn with TikZ.}
   \label{Fig:_diagram1}
\end{figure}
\end{verbatim}\txcol
\end{minipage}
\begin{minipage}{0.39\linewidth}
\brocol\begin{verbatim}
\begin{align*}
   \int dx\; f(x) =		
   \alpha
%\tikzset{external/force remake}
   \begin{tikzpicture}
		
      ... TikZ commands ...

   \end{tikzpicture}
\end{align*}
\end{verbatim}\txcol
\end{minipage}
%
\subsection{Layers}
\label{Apx_tikzbasics_layers}
\noindent
In general, objects that are drawn later in the code appear 
on top of objects that are drawn earlier (if they overlap).
If we need to declare explicitly which objects
we want to appear on top of other objects,
then we can use layers.

Layers are first declared with \brocol\verb$\pgfdeclarelayer$\txcol
as in the following example, and then their order is set up
from backmost to foremost with \brocol\verb$\pgfsetlayers$\txcol.
The layer \brocol\verb$main$\txcol is predefined and does not need 
to be declared, but it must always be included in the order setup,
for example:
\brocol\begin{verbatim}
\pgfdeclarelayer{background}
\pgfdeclarelayer{foreground}
\pgfsetlayers{background,main,foreground}
\end{verbatim}\txcol
All drawing commands by default are placed on the layer 
\brocol\verb$main$\txcol.
For placing them on other layers, they must be enclosed in the 
\brocol\verb$\pgfonlayer$\txcol environment as in the following example
(the \brocol\verb$opacity$\txcol option is optional).\\

\noindent
\begin{minipage}{0.83\linewidth}
\brocol\begin{verbatim}
  \begin{pgfonlayer}{foreground}
     \filldraw[fill=Orange, draw=Red, opacity=0.7] (1,0) circle [radius=6mm];
  \end{pgfonlayer}
  \filldraw[fill=Blue, draw=black] (1,0) circle [radius=9mm];
\end{verbatim}\txcol
\end{minipage}
\begin{minipage}{0.16\linewidth}
\tikzsetnextfilename{tikz_layers}
\begin{tikzpicture}
  \pgfdeclarelayer{foreground}
  \pgfsetlayers{main,foreground}
  \begin{pgfonlayer}{foreground}
     \filldraw[fill=Orange,draw=Red, opacity=0.7] (1,0) circle [radius=6mm];
  \end{pgfonlayer}
  \filldraw[fill=Yellow,draw=black] (0,0) circle [radius=8mm];
\end{tikzpicture}
\end{minipage}
%
\subsection{Drawing objects}
\label{Apx_tikzbasics_drawing}
\noindent
All drawing commands must end with a \;;\; (semicolon).
%
\subsubsection{Text and formulas}
\noindent
Plain text and LaTex math formulas can be inserted anywhere 
in the graphic with TikZ' \brocol\verb$node$\txcol command.
The text is centered at the node's coordinates:\\

\noindent
\begin{minipage}{0.83\linewidth}
\brocol\begin{verbatim}
  \filldraw[fill=Yellow, draw=Red] (0,0) circle [radius=1cm];
  \node at (-0.8,-0.3)  {\textcolor{Blue}{\bfseries Yes! $\alpha^2$}};
\end{verbatim}\txcol
\end{minipage}
\begin{minipage}{0.16\linewidth}
\tikzsetnextfilename{tikz_text_math}
\begin{tikzpicture}
  \filldraw[fill=Yellow, draw=Red] (0,0) circle [radius=1cm];
  \node at (-0.8,-0.3)  {\textcolor{Blue}{\bfseries Yes! $\alpha^2$}};
\end{tikzpicture}
\end{minipage}
%
%
\subsubsection{Lines: straight and smooth}
\noindent
A straight line from node (a) to node (b) is drawn by\\

\noindent
\begin{minipage}{0.83\linewidth}
\brocol\begin{verbatim}
  \draw [Red, line width = 2pt] (a1) to (a2);
  \draw (b1) to [Blue] (b2);
\end{verbatim}\txcol
\end{minipage}
\begin{minipage}{0.16\linewidth}
\tikzsetnextfilename{tikz_lines_straight}
\begin{tikzpicture}
  \node (a1) at (0,0) {};
  \node (a2) at (1,0.5) {};
  \node (b1) at (2,0) {};
  \node (b2) at (3,0.5) {};
  \draw [Red, line width = 2pt] (a1) to (a2);
  \draw (b1) to [Blue] (b2);
\end{tikzpicture}
\end{minipage}
A smooth curve which leaves (a) at an angle of 45 degrees 
and enters (b) at 180 degrees is drawn by\\

\noindent
\begin{minipage}{0.83\linewidth}
\brocol\begin{verbatim}
  \draw [Green, line width=1mm] (a1) to [out=45, in=180] (a2);
\end{verbatim}\txcol
\end{minipage}
\begin{minipage}{0.16\linewidth}
\tikzsetnextfilename{tikz_lines_in_out}
\begin{tikzpicture}
  \node (a1) at (0,0) {};
  \node (a2) at (1,0.5) {};
  \draw [Green, line width=1mm] (a1) to [out=45, in=180] (a2);
\end{tikzpicture}
\end{minipage}
\\
\noindent
The angles are with respect to the coordinate system, that is: 
0 degrees points right, 90 up, 180 left and 270 down.
%
%
\subsubsection{Lines with arrows}
Lines can be equipped with arrows by specifying this
in the options of \brocol\verb$\draw$\txcol as inthe examples below.
The most basic arrow styles are:
\brocol\verb$>, Latex, Stealth, Circle, Bar$\txcol.
Note the extra curly braces!\\

\noindent
\begin{minipage}{0.87\linewidth}
\brocol\begin{verbatim}
\draw [->] (a1) to (a2);
\draw [<->,Red, line width = 1pt] (b1) to [out=45, in=135] (b2);
\draw [{Circle[width=5pt,length=5pt]}-{Latex[length=8pt,width=6pt]}] (c1) to (c2);
\draw [{Bar[Red,width=8pt]}-{Stealth[length=8pt, width=6pt]}] (d1) to (d2);
\end{verbatim}\txcol
\end{minipage}
\begin{minipage}{0.12\linewidth}
\tikzsetnextfilename{tikz_arrows}
\begin{tikzpicture}
  \node (a1) at (0,0) {};
  \node (a2) at (2,0) {};
  \node (b1) at (0,-0.7) {};
  \node (b2) at (2,-0.7) {};
  \node (c1) at (0,-1) {};
  \node (c2) at (2,-1) {};
  \node (d1) at (0,-1.5) {};
  \node (d2) at (2,-1.5) {};
  \draw [->] (a1) to (a2);
  \draw [<->,Red, line width = 1pt] (b1) to [out=45, in=135] (b2);
  \draw [{Circle[width=5pt,length=5pt]}-{Latex[length=8pt, width=6pt]}] (c1) to (c2);
  \draw [{Bar[Red,width=8pt]}-{Stealth[length=8pt, width=6pt]}] (d1) to (d2);
\end{tikzpicture}
\end{minipage}
%
\subsection{Externalizing graphics}
\label{Apx_tikzbasics_externalize}
\noindent
Externalizing means that TikZ creates a separate PDF file
for each \brocol\verb$tikzpicture$\txcol environment
which it encounters in the document.
That is, when a \brocol\verb$tikzpicture$\txcol is encountered,
TikZ checks whether there is already a corresponding PDF,
and if so, then it includes this PDF with \brocol\verb$\includegraphics$\txcol
and ignores the code in the \brocol\verb$tikzpicture$\txcol,
else it creates the corresponding PDF such that it can be used
in the next LaTex run.
Using the external PDFs can speed up the LaTex runs significantly.
TikZ also takes care of vertically positioning the externalized PDFs
according to the \brocol\verb$baseline$\txcol option
when it is used with \brocol\verb$tikzpicture$\txcol
(this is what the .dpth files are for).
TikZ recognizes automatically when the TikZ code of a graphic has been modified,
and generates fresh PDFs \emph{only} for these modified graphics in the next LaTex run.

In order to activate externalizing, we need to load the library with
\brocol\verb$\usetikzlibrary{external}$\txcol
(which both TikZ-Feynman and FeynHand do automatically)
and put a \brocol\verb$\tikzexternalize[prefix=graphics/tikz/]$\txcol
somewhere above the \brocol\verb$\begin{document}$\txcol command.
Here, we use the \brocol\verb$prefix$\txcol option
for telling TikZ that we want the external files to be created
in the subdirectory \brocol\verb$graphics/tikz/$\txcol of the current dirctory.
It may come in handy to have the TikZ-generated files
in a subfolder separate from the other external graphic files.
For example, in order to assure that the final document contains
the final version of all graphics, we can force TikZ to refresh them
by deleting all external PDFs created by TikZ.
Having these in the separate subfolder avoids deleting graphics from other sources.
Alternatively, we can include \brocol\verb$\tikzset{external/force remake}$\txcol
in the document's preamble to force refresh all external PDFs.
Including \brocol\verb$\tikzset{external/force remake}$\txcol
in a local Tex group force refreshes all external PDFs in this group
(e.g.~an equation, figure or minipage environment).

\begin{figure}
	\centering
	\includegraphics[width=0.75\linewidth]{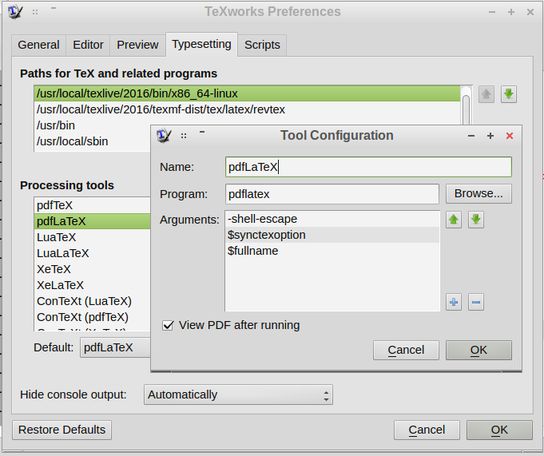}
	\caption{TexWorks screenshot.}
	\label{Fig:_shell_escape}
\end{figure}

We also have to tell LaTex to use shell escaping
in order to enable it to create the external files.
TexLive ships with the TexWorks editor, in which this can be achieved
by going to the Edit menu and selecting Preferences.
Next choose the Typesetting rider, select PdfLatex in
the Processing Tools window, and add the argument
-shell-escape by clicking on the + button.
Then, move it to the top of the list by clicking on the up-arrow button.
This should look somewhat like Figure~\ref{Fig:_shell_escape}.

TikZ creates filenames automatically for the externalized PDFs.
For later use of these, it is often preferrable
to specify these filenames by hand with the
\brocol\verb$\tikzsetnextfilename$\txcol command,
which needs to be placed right above
the \brocol\verb$\begin{tikzpicture}$\txcol
of the corresponding graphic.
For example, \brocol\verb$\tikzsetnextfilename{scatter}$\txcol
tells TikZ to store the next graphic as \brocol\verb$scatter.pdf$\txcol

Apart from speeding up LaTex runs, the external files can also be used
for submissions to journals which do not support TikZ.
One way of doing this works without modifying the graphics' code:
Instead of \brocol\verb$\usepackage{tikz}$\txcol
respectively \brocol\verb$\usepackage{feynhand}$\txcol
we put \brocol\verb$\usepackage{tikzexternal}$\txcol
and include the file \brocol\verb$tikzexternal.sty$\txcol
in our submissions. It basically tells LaTex to ignore
the code in the \brocol\verb$tikzpicture$\txcol environments
and to use the external PDFs instead.
If this does not work, then another way is to replace the figures' 
\brocol\verb$tikzpicture$\txcol environments by
\brocol\verb$\includegraphics{tikz/filename.pdf}$\txcol 
and the equations' \brocol\verb$tikzpicture$\txcol environments
by \brocol\verb$\vcenterbox{\includegraphics{tikz/filename.pdf}}$\txcol 
commands, having defined in the preamble:
\brocol
\begin{verbatim}
   \graphicspath{{graphics/}}     % mind the double brackets!
   \newcommand{\vcenterbox}[2][0.5]{\raisebox{-#1\height}{#2}}
   % \vcenterbox centers the box's content (argument #2) vertically on the baseline.
   % The optional argument #1 is for fine adjustment:
   %      values between 0 and 0.5 raise the box's center above the baseline,
   %      values between 0.5 and 1 push the box's center below the baseline.
\end{verbatim}
\txcol
%
%
{}
%
%
%
\end{document}